\documentstyle[12pt]{article}
\begin{document}
%
\pagestyle{empty}
\begin{center}
{\large\bf QCD Predictions for $g_1^p$ at Small $x$ 
Incorporating Double $ln^2(1/x)$   
 Resummation \footnote{ To appear in the Proceedings of the 1997 Workshop on 'Physics with Polarized Protons at
HERA', eds. A.De Roeck and T.Gehrmann }}
\vspace{1.1cm}\\ 
         {\sc J.~Kwieci\'nski}, 
         {\sc B.~Ziaja}\\ 
\vspace{0.3cm}
{\it Department of Theoretical Physics, \\
H.~Niewodnicza\'nski Institute of Nuclear Physics, 
Cracow, Poland}
\end{center}
\vspace{1.1cm}
\begin{abstract}  
The proton spin dependent 
structure function  $g_1^p$ 
is  analysed using the unified scheme 
incorporating both Altarelli-Parisi evolution and the double $ln^2(1/x)$ 
effects at low $x$. The latter are found to be very important in the region 
of $x$ which can possibly be probed at HERA.  Predictions for the polarized 
gluon distribution $\Delta G(x)$ at low $x$ are also given.   
\end{abstract}
\vspace{1.1cm}
It has recently been pointed out that the spin dependent structure function 
$g_1$ at low $x$ is dominated  by the {\it double} logarithmic $ln^2(1/x)$ 
contributions 
i.e. by those terms of the perturbative expansion which correspond to 
the powers of $ln^2(1/x)$ at each order of the expansion 
\cite{BARTNS,BARTS}.  Those contributions go {\it beyond} 
the LO or NLO order QCD evolution of polarised parton densities \cite{AP} and in order to take 
them into account one has to 
include the resummed double $ln^2(1/x)$ terms in the coefficient 
and splitting functions \cite{RG}. In ref. \cite{BBJK} we have developed an 
alternative approach based on unitegrated distributions which for simplicity 
was formulated for the non-singlet distributions.  In this note we wish to 
present the preliminary results for the singlet structure functions  
concentrating on the proton polarized structure function $g_1^p(x,Q^2)$. Our main 
aim is to study $g_1^p(x,Q^2)$ in the 
region of the small values of $x$ which can possibly be probed at HERA 
\cite{ALBERT} and confront obtained results with those obtained from the LO 
Altarelli-Parisi evolution.  We will also show similar results for the polarised 
gluon distribution $\Delta G(x,Q^2)$.\\

The dominant contribution to the double $ln^2(1/x)$ resummation comes from 
the ladder diagrams with quark and gluon exchanges along the ladder.  
In what follows we shall neglect for simplicity 
possible 
non-ladder bremsstrahlung terms which are relatively unimportant 
\cite{BARTNS,BARTS}.\\

It is convenient to introduce the unintegrated (spin dependent) parton 
distributions 
$f_i(x^{\prime},k^2)$ ($i=u_{v},d_{v},\bar u,\bar d,\bar s,g$) where $k^2$ 
is the transverse momentum squared of the parton $i$ and $x^{\prime}$ the 
longitudinal momentum fraction of the parent proton carried by a parton.  
The conventional (integrated) distributions $\Delta p_i(x,Q^2)$ are related in the 
following way  to the unintegrated distributions $f_i(x^{\prime},k^2)$: 
\begin{equation}
 \Delta p_i(x,Q^2)=\Delta p_i^{(0)}(x)+
 \int_{k_0^2}^{W^2}{dk^2\over k^2}f_i(x^{\prime}=
x(1+{k^2\over Q^2}),k^2)
\label{dpi}
\end{equation}
where $\Delta p_i^{(0)}(x)$ is the nonperturbative part of the 
of the distributions  
and 
\begin{equation}
W^2=Q^2({1\over x}-1)
\label{w2}
\end{equation} 
where as usual $x$ is the Bjorken parameter and $Q^2=-q^2$ with $q$ denoting the 
four momentum transfer between leptons.  
The parameter $k_0^2$ is the infrared cut-off which will be set equal 
to 1 GeV$^2$. 
The origin of the nonperturbative part $\Delta p_i^{(0)}(x)$ can be viewed
upon as originating from the non-perturbative region  $k^2<
k_0^2$, i.e. 
\begin{equation}
\Delta p_i^{(0)}(x)= \int_{0}^{k_0^2}{dk^2\over k^2}f_i(x,k^2)    
\label{gint0}
\end{equation} 
The spin dependent structure function $g_1^p(x,Q^2)$ of the proton is 
related in a standard 
way to the (integrated) parton distributions:
\begin{equation}
g_1^p(x,Q^2)={1\over 2}\left[{4\over 9}(\Delta u_v(x,Q^2) + 2\Delta \bar u (x,Q^2))+
{1\over 9}(\Delta d_v(x,Q^2) + 2\Delta \bar u(x,Q^2) +  2\Delta \bar s(x,Q^2))
\right]
\label{gp1}
\end{equation}                           
where $\Delta u_v(x,Q^2) = \Delta p_{u_v}(x,Q^2)$ etc.  
We assume $\Delta \bar u =
\Delta \bar d$ and confine ourselves to the number of flavours $N_F$ equal to 
three.\\

The valence quarks distributions and asymmetric part of the sea 
\begin{equation} 
f_{US}(x^{\prime},k^2)= f_{\bar u}(x^{\prime},k^2)-f_{\bar s}(x^{\prime},k^2)
\label{us}
\end{equation}  
will correspond to  ladder diagrams with quark exchange along the ladder.  
The singlet distributions 
\begin{equation}
f_{S}(x^{\prime},k^2) = f_{u_v}(x^{\prime},k^2)+f_{d_v}(x^{\prime},k^2)+
4f_{\bar u}(x^{\prime},k^2)+ 
2f_{\bar s}(x^{\prime},k^2)
\label{sns}
\end{equation}
and the gluon distributions $f_g(x^{\prime},k^2)$ will correspond to ladder 
diagrams with both quark (antiquark) and gluon exchanges along the ladder. \\
   
The sum of double logarithmic $ln^2(1/x)$ 
terms corresponding to ladder diagrams is generated by the following 
integral equations: 
\begin{equation} 
f_{k}(x^{\prime},k^2)=f^{(0)}_{k}(x^{\prime},k^2) + 
{\alpha_s\over 2 \pi} \Delta P_{qq}(0)   
\int_{x^{\prime}}^1 {dz\over z} 
\int_{k_0^2}^{k^2/z}
{dk^{\prime 2}\over k^{\prime 2}} 
f_{k}\left({x^{\prime}\over z},k^{\prime 2}\right)
\label{dlxns}
\end{equation}
($k= u_v, d_v, US$)

$$ 
f_{S}(x^{\prime},k^2)=f^{(0)}_{S}(x^{\prime},k^2) +
$$ 
$$ 
{\alpha_s\over 2 \pi}    
\int_{x^{\prime}}^1 {dz\over z} 
\int_{k_0^2}^{k^2/z}
{dk^{\prime 2}\over k^{\prime 2}} 
\left[\Delta P_{qq}(0)
f_{S}\left({x^{\prime}\over z},k^{\prime 2}\right)+
\Delta P_{qg}(0)
f_{g}\left({x^{\prime}\over z},k^{\prime 2}\right)\right]
$$

$$ 
f_{g}(x^{\prime},k^2)=f^{(0)}_{g}(x^{\prime},k^2) + 
$$
\begin{equation}
{\alpha_s\over 2 \pi}    
\int_{x^{\prime}}^1 {dz\over z} 
\int_{k_0^2}^{k^2/z}
{dk^{\prime 2}\over k^{\prime 2}} 
\left[\Delta P_{gq}(0)
f_{S}\left({x^{\prime}\over z},k^{\prime 2}\right)+
\Delta P_{gg}(0)
f_{g}\left({x^{\prime}\over z},k^{\prime 2}\right)\right]
\label{dlxsg}
\end{equation} 
with $\Delta P_{ij}(0) = \Delta P_{ij}(z=0)$ where $\Delta P_{ij}(z)$ 
denote the LO splitting functions describing evolution 
of spin dependent  parton densities and the inhomogeneous terms 
$f^{(0)}(x,^{\prime},k^2)$ will be specified later. To be precise we have: 
$$
\Delta P_{qq}(0) = {4\over 3} 
$$
$$
\Delta P_{qg}(0)= - N_F 
$$
$$
\Delta P_{gq}(0) = {8\over 3} 
$$
\begin{equation}
\Delta P_{gg}(0)= 12 
\label{dpij}
\end{equation}

Equation (\ref{dlxns}) is similar to the corresponding equation in QED 
describing the double logarithmic resummation generated by ladder diagrams with 
fermion exchange \cite{QED}. The problem of double logarithmic 
asymptotics in QCD in the non-singlet channels was also discussed in ref  
\cite{QCD}.\\

Equations (\ref{dlxns}, \ref{dlxsg}) generate singular power behaviour of the 
spin dependent parton distributions and of the spin dependent 
structure functions $g_1$  at small  $x$ i.e.  
$$ 
g_1^{NS}(x,Q^2) \sim x^{-\lambda_{NS}} 
$$ 
$$
g_1^{S}(x,Q^2) \sim x^ {-\lambda_{S}}
$$
\begin{equation}
\Delta G(x,Q^2) \sim x^ {-\lambda_{S}}
\label{power}
\end{equation}
where $g_1^{NS}=g_1^p-g_1^n$ and $g_1^{S}=g_1^p+g_1^n$ respectively and 
$\Delta G$ is the spin dependent gluon distribution.  This behaviour reflects 
 similar small $x^{\prime}$ behaviour of the unintegrated distributions.  
Exponents $\lambda_{NS,S}$ are given by the following formulas:  
$$
\lambda_{NS} = 2 \sqrt{\left[{\alpha_s \over 2\pi} \Delta P_{qq}(0)\right]}
$$
\begin{equation}
\lambda_{S} = 2 \sqrt{\left[{\alpha_s \over 2\pi} \gamma^+\right]}
\label{lambda}
\end{equation}
where 
\begin{equation}
\gamma^+ = {\Delta P_{qq}(0) +\Delta P_{gg}(0) + 
\sqrt{(\Delta P_{qq}(0) -\Delta P_{gg}(0))^2 + 
4\Delta P_{qg}(0)\Delta P_{gq}(0)}\over 2}
\label{lplus}
\end{equation}
Both  equations (\ref{dlxns},\ref{dlxsg}) and the exponents $\lambda_{NS,S}$ 
acquire additional contributions from the non-ladder diagrams which are however 
relatively small \cite{BARTNS,BARTS}.  The power-like behaviour 
(\ref{lambda}) remains 
the leading small $x$ behaviour of the structure functions provided that 
their non-perturbative parts are less singular.  This takes place 
if the latter are assumed to have the Regge pole like behaviour with the 
corresponding intercept(s) being near $0$. \\

Following ref. \cite{BBJK} we extend  equations (\ref{dlxns},\ref{dlxsg}) and 
add to their right hand side the contributions coming from the 
remaining parts of the splitting functions $\Delta P_{ij}(z)$.  
We also allow the coupling $\alpha_s$ to run setting $k^2$ as the relevant 
scale. In this way we obtain unified system of equations which contain 
both the complete LO Altarelli-Parisi evolution and the double logarithmic 
$ln^2(1/x)$ effects at low $x$.  
The corresponding system of equations reads:
$$ 
f_{k}(x^{\prime},k^2)=f^{(0)}_{k}(x^{\prime},k^2) + 
{\alpha_s(k^2)\over 2 \pi}{4\over 3}   
\int_{x^{\prime}}^1 {dz\over z} 
\int_{k_0^2}^{k^2/z}
{dk^{\prime 2}\over k^{\prime 2}} 
f_{k}\left({x^{\prime}\over z},k^{\prime 2}\right)+
$$
$$
{\alpha_s(k^2)\over 2\pi}\int_{k_0^2}^{k^2}{dk^{\prime 2}\over
k^{\prime 2}}{4\over 3}\int _{x^{\prime}}^1
{dz\over z} {(z+z^2)f_k({x^{\prime}\over z},k^{\prime 2})-
2zf_{k}(x^{\prime},k^{\prime 2})\over 1-z}+
$$
\begin{equation}
{\alpha_s(k^2)\over 2\pi}\int_{k_0^2}^{k^2}{dk^{\prime 2}\over
k^{\prime 2}}\left[2 +
{8\over 3} ln(1-x^{\prime})\right]f_{k}(x^{\prime},k^{\prime 2})
\label{unifns}
\end{equation}
($k=u_v, d_v, US$),  

$$ 
f_{S}(x^{\prime},k^2)=f^{(0)}_{S}(x^{\prime},k^2) +{\alpha_s(k^2)\over 2 \pi}    
\int_{x^{\prime}}^1 {dz\over z} 
\int_{k_0^2}^{k^2/z}
{dk^{\prime 2}\over k^{\prime 2}} 
{4\over 3} 
f_{S}\left({x^{\prime}\over z},k^{\prime 2}\right) +
$$ 
$$ 
{\alpha_s(k^2)\over 2 \pi}\int_{k_0^2}^{k^2}{dk^{\prime 2}\over
k^{\prime 2}}{4\over 3}\int _{x^{\prime}}^1
{dz\over z} {(z+z^2)f_S({x^{\prime}\over z},k^{\prime 2})-
2zf_S(x^{\prime},k^{\prime 2})\over 1-z}+
$$
$$
{\alpha_s(k^2)\over 2 \pi}\int_{k_0^2}^{k^2}{dk^{\prime 2}\over
k^{\prime 2}}\left[ 2 +
{8\over 3} ln(1-x^{\prime})\right]
f_{S}(x^{\prime},k^{\prime 2}) + 
$$
$$ 
{\alpha_s(k^2)\over 2 \pi}N_F\left[-\int_{x^{\prime}}^1 {dz\over z} 
\int_{k_0^2}^{k^2/z}
{dk^{\prime 2}\over k^{\prime 2}}f_{g}
\left({x^{\prime}\over z},k^{\prime 2}\right) + \int_{k_0^2}^{k^2}{dk^{\prime 2}\over
k^{\prime 2}}\int _{x^{\prime}}^1
{dz\over z} 2z f_g({x^{\prime}\over z},k^{\prime 2})\right] 
$$
 
$$ 
f_{g}(x^{\prime},k^2)=f^{(0)}_{g}(x^{\prime},k^2) + {\alpha_s(k^2)\over 2 \pi}    
\int_{x^{\prime}}^1 {dz\over z} 
\int_{k_0^2}^{k^2/z}
{dk^{\prime 2}\over k^{\prime 2}} 
{8\over 3}
f_{S}\left({x^{\prime}\over z},k^{\prime 2}\right) + 
$$
$$
{\alpha_s(k^2)\over 2 \pi}\left[     
\int_{k_0^2}^{k^2}
{dk^{\prime 2}\over k^{\prime 2}}\int_{x^{\prime}}^1 {dz\over z}  
(-{4\over 3})zf_{S}
\left({x^{\prime}\over z},k^{\prime 2}\right) + 
12   
\int_{x^{\prime}}^1 {dz\over z} 
\int_{k_0^2}^{k^2/z}
{dk^{\prime 2}\over k^{\prime 2}} 
 f_{g}\left({x^{\prime}\over z},k^{\prime 2}\right)\right] +
$$ 
$$
{\alpha_s(k^2)\over 2 \pi}    
\int_{k_0^2}^{k^2}
{dk^{\prime 2}\over k^{\prime 2}} \int_{x^{\prime}}^1 {dz\over z} 
6z\left[{f_{g}
\left({x^{\prime}\over z},k^{\prime 2}\right)- f_{g}
(x^{\prime},k^{\prime 2})\over 1-z} -2f_{g}
\left({x^{\prime}\over z},k^{\prime 2}\right)\right]+
$$
\begin{equation}
{\alpha_s(k^2)\over 2 \pi}    
\int_{k_0^2}^{k^2}
{dk^{\prime 2}\over k^{\prime 2}}\left[ {11\over 2} -{N_F\over 3}
 + 6 ln(1-x^{\prime})\right]f_{g}
(x^{\prime},k^{\prime 2})
\label{unifsg}
\end{equation}

The inhomogeneous terms $f_i^{(0)}(x^{\prime},k^2)$ are expressed in terms of the input (integrated) 
parton distributions and are the same as in the case of the LO Altarelli 
Parisi evolution \cite{BBJK}: 
$$
f_k^{(0)}(x^{\prime},k^2)={\alpha_s(k^2)\over 2 \pi}{4\over 3}
\int _{x^{\prime}}^1
{dz\over z} {(1+z^2)\Delta p_k^{(0)}({x^{\prime}\over z})-
2z\Delta p_k^{(0)}(x^{\prime})\over 1-z}+
$$
\begin{equation}
{\alpha_s(k^2)\over 2 \pi}\left[2 +{8\over 3} 
ln(1-x^{\prime})\right]\Delta p_k^{(0)}(x^{\prime}) 
\label{fns0}
\end{equation}
($k=u_v,d_v,US$), 

$$
f_S^{(0)}(x^{\prime},k^2) = {\alpha_s(k^2)\over 2 \pi}{4\over 3}
\int _{x^{\prime}}^1
{dz\over z} {(1+z^2)\Delta p_S^{(0)}({x^{\prime}\over z})-
2z\Delta p_S^{(0)}(x^{\prime})\over 1-z}+
$$
$$
{\alpha_s(k^2)\over 2 \pi}\left[(2 +{8\over 3} ln(1-x^{\prime}))
\Delta p_S^{(0)}(x^{\prime})+
N_F\int _{x^{\prime}}^1{dz\over z} 
(1-2z)\Delta p_g^{(0)}({x^{\prime}\over z})\right] 
$$

$$
f_g^{(0)}(x^{\prime},k^2) ={\alpha_s(k^2)\over 2 \pi}\left[{4\over 3}
\int _{x^{\prime}}^1{dz\over z}(2-z)\Delta p_S^{(0)}({x^{\prime}\over z}) + ({11\over 2} -{N_F\over 3}
 + 6 ln(1-x^{\prime}))\Delta p_{g}^{(0)}
(x^{\prime})\right]+ 
$$
\begin{equation} 
{\alpha_s(k^2)\over 2 \pi}6    
\int_{x^{\prime}}^1 {dz\over z}\left[ 
{\Delta p_{g}^{(0)}
({x^{\prime}\over z})- z\Delta p_{g}^{(0)}
(x^{\prime})\over 1-z} +(1-2z)\Delta p_{g}^{(0)}
({x^{\prime}\over z})\right]
\label{fsg0}
\end{equation}
 
Equations (\ref{unifns},\ref{unifsg}) together with (\ref{fns0},\ref{fsg0}) and 
(\ref{dpi}) reduce to the LO Altarelli-Parisi evolution equations with the 
starting (integrated) distributions $\Delta p_i^{0}(x)$  
after we set the upper 
integration limit over $dk^{\prime 2}$ equal to $k^2$ in all terms in 
equations (\ref{unifns},\ref{unifsg}) and if we set $Q^2$ in place of $W^2$ as 
the upper integration limit in the integral in eq. (\ref{dpi}).

We solved  equations (\ref{unifns},\ref{unifsg})  assuming the following simple parametrisation of the 
input distributions: 

\begin{equation}
\Delta p_i^{(0)}(x)=N_i (1-x)^{\eta_i}
\label{dpi0}
\end{equation} 
where we set $\eta_{u_v}=\eta_{d_v}=3,$  $\eta_{\bar u} = \eta_{\bar s} = 7 $ 
and $\eta_g=5$.  The normalisation constants $N_i$ were determined 
 by imposing the Bjorken sum-rule for $\Delta u_v^{(0)}-\Delta d_v^{(0)}$ 
 and requiring that the first moments of 
all other distributions are the same as those determined from  the recent 
QCD analysis \cite{STRATMAN}. All distributions $\Delta p_i^{(0)}(x)$ 
behave as $x^0$ in the limit $x\rightarrow 0$ that corresponds to the implicit 
assumption that the Regge poles which correspond to axial vector mesons, which 
should control the small $x$ behaviour of $g_1$ \cite{REGGE} have their intercept equal 
to $0$.  We checked that the parametrisation (\ref{dpi0}) combined with 
equations 
(\ref{dpi},\ref{gp1},\ref{unifns},\ref{unifsg}) gives  
reasonable description of the recent SMC data on $g_1^p(x,Q^2)$ \cite{SMC}.\\

In Fig.1 we show $g_1^p(x,Q^2)$ for $Q^2= 10 GeV^2$ in the small $x$ 
region which can  possibly 
be probed at HERA.  We show predictions based on  the equations 
(\ref{unifns},\ref{unifsg},\ref{dpi}) and confront them with the expectations 
which follow 
from solving the LO Altarelli-Parisi evolution equations with the input 
distributions at $Q_0^2=1 GeV^2$ given by equation(s) (\ref{dpi0}).    
We also show in this Figure the "experimental" points which were obtained 
from the extrapolations based on the NLO QCD analysis together with estimated 
statistical errors \cite{ALBERT}. 
We see that the structure function $g_1^p(x,Q^2)$ which contains effects 
of the double $ln^2(1/x)$ resummation begins to differ from that calculated within 
the LO Altarelli Parisi equations already for $x \sim 10^{-3}$.  
It is however comparable to the structure function obtained from the 
NLO analysis for $x>10^{-4}$ which is indicated by the "experimental" points. 
This is presumably partially an artifact of the difference in the input distributions but 
it also reflects the fact that the NLO approximation contains the 
first two terms of the double $ln^2(1/x)$ resummation in the corresponding 
splitting and coefficient functions. It can also be seen from Fig.1 that  the 
(complete) double $ln^2(1/x)$ 
resummation generates the structure function which is significantly steeper than 
that obtained from the  NLO QCD analysis and the difference between those two extrapolations 
becomes significant for $x < 10^{-4}$.\\
    
In Fig. 2 we show our predictions 
for the polarized gluon distribution $\Delta G(x,Q^2) = \Delta p_g(x,Q^2)$ for 
$Q^2=10 GeV^2$ and confront it with the polarized gluon distribution obtained 
from the LO Altarelli-Parisi equations.  The gluon distributions exhibits characteristic 
$x^{-\lambda_S}$ behaviour with $\lambda_S \sim 1$.  Similar behaviour is 
exhibited by the structure function $g_1^p(x,Q^2)$ itself.\\

To sum up we have presented results of the analysis of the "unified" equations 
which contain the LO Altarelli Parisi evolution and the double $ln^2(1/x)$ effects 
at low $x$.  As the first approximation we considered  those double 
$ln^2(1/x)$ effects which are generated by ladder diagrams.  
The double logarithmic effects were found to be very important and they should 
 in principle be visible in possible HERA measurements (cf. Fig. 1). \\
                                                                                                                                                                                                                                                                                                                                                                                                                                                                                                                                        
\section*{Acknowledgments}
We thank Barbara Bade\l{}ek, Albert DeRoeck and Marco Stratmann for useful 
and illuminating discussions.  We are also grateful to Thomas Gehrmann 
and Albert DeRoeck for their encouragement to write up this contribution.  
This research has been supported in part by the Polish Committee for Scientific 
Resarch grants 2 P03B 184 10 and 2 P03B 89 13. 

\section*{Figure Captions}
Fig.1 The structure function $g_1^p(x,Q^2)$ for $Q^2=10 GeV^2$ plotted as 
the function 
of $x$.  Solid line represents this structure function with the double 
$ln^2(1/x)$ terms included and the dashed line corresponds to $g_1^p$ obtained  
from the LO Altarelli-Parisi equations  The "experimental" points are based on 
the NLO QCD predictions with the statistical errors expected at HERA 
\cite{ALBERT}.\\

Fig.2  The spin dependent gluon distribution $\Delta G(x,Q^2)$  
for $Q^2=10 GeV^2$ plotted as the function 
of $x$.  Solid line represents $\Delta G(x,Q^2)$   with the double 
$ln^2(1/x)$ terms included and the dashed line corresponds to 
the $\Delta G(x,Q^2)$ 
obtained from the LO Altarelli-Parisi equations.     
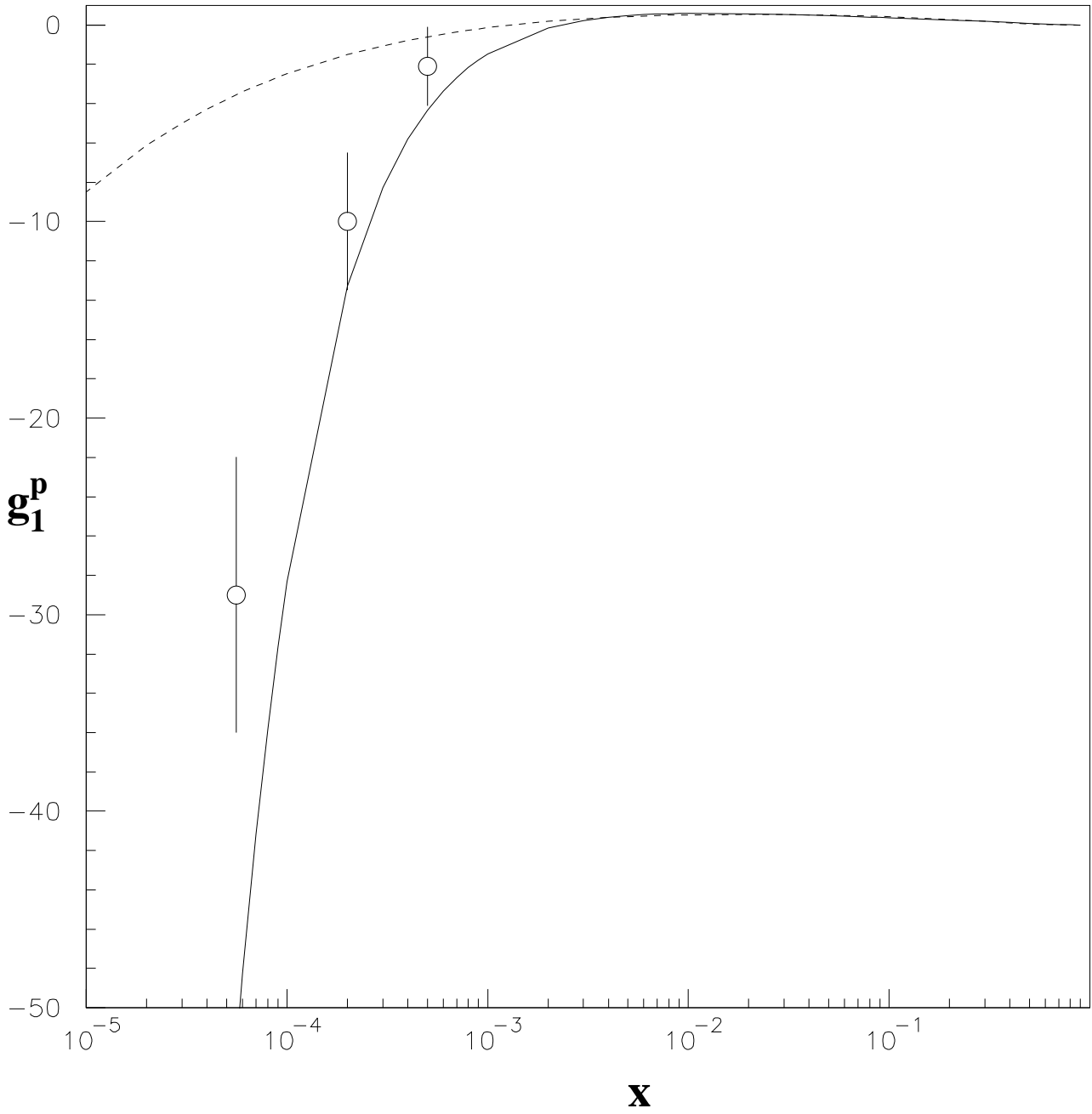
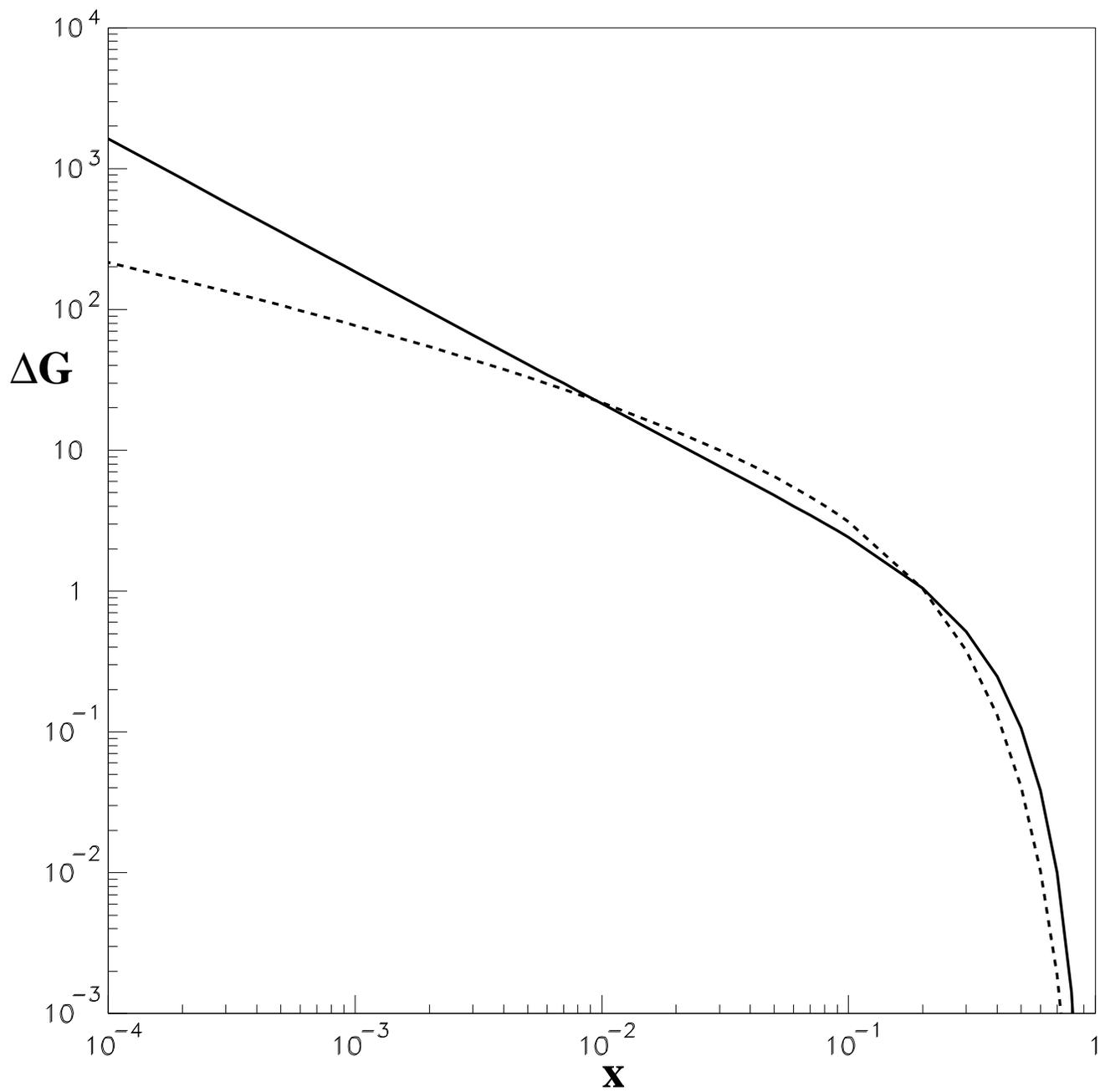
\end{document}